\documentclass[showpacs,amsmath,amssymb,PRB,superscriptaddress,,reprint]{revtex4}
\usepackage{graphicx}
\usepackage{multirow}
\usepackage{textcomp}
\usepackage{color} 

\usepackage[colorlinks,plainpages=false,linkcolor=blue,urlcolor=blue,citecolor=blue,pdfpagemode=UseNone,pdfstartview=FitBH]{hyperref}
\usepackage{latexsym}        
\usepackage{amssymb}
\usepackage{amsmath}
\usepackage{amsfonts}

\begin{document}
\title[Temperature dependent local structure of SmFe$_{1-x}$Ru$_x$AsO$_{0.85}$F$_{0.15}$]{Temperature dependent local atomic displacements in Ru substituted SmFe$_{1-x}$Ru$_x$AsO$_{0.85}$F$_{0.15}$ superconductors}

\author{Boby Joseph} 
\affiliation{Dipartimento di Fisica, Universit\`{a} di Roma ``La
Sapienza", P. le Aldo Moro 2, 00185 Roma, Italy} 
\author{Antonella Iadecola}
\affiliation{Dipartimento di Fisica, Universit\`{a} di Roma ``La
Sapienza", P. le Aldo Moro 2, 00185 Roma, Italy} 
\affiliation{Elettra, Sincrotrone Trieste, Strada Statale 14, Km 
163.5, Basovizza, Trieste, Italy}
\author{Laura Simonelli}
\address{European Synchrotron Radiation Facility, 6 RUE Jules
Horowitz BP 220 38043 Grenoble Cedex 9 France}
\author{Laura Maugeri} 
\affiliation{Dipartimento di Fisica, Universit\`{a} di Roma ``La
Sapienza", P. le Aldo Moro 2, 00185 Roma, Italy} 
\author{Alberto Martinelli}
\affiliation{CNR-SPIN and Universit\`{a} di Genova, via Dodecaneso 33, 16146 Genova, Italy}
\author{Andrea Palenzona}
\affiliation{CNR-SPIN and Universit\`{a} di Genova, via Dodecaneso 33, 16146 Genova, Italy}
\author{Marina Putti} 
\affiliation{CNR-SPIN and Universit\`{a} di Genova, via Dodecaneso 33, 16146 Genova, Italy}
\author{Naurang L. Saini}
\affiliation{Dipartimento di Fisica, Universit\`{a} di Roma ``La Sapienza", P. le Aldo Moro 2, 00185 Roma, Italy} 

\begin{abstract}
Local structure of SmFe$_{1-x}$Ru$_x$AsO$_{0.85}$F$_{0.15}$ ($x$ = 0.0,
0.05, 0.25 and 0.5) superconductors has been investigated by
temperature dependent As $K$-edge extended x-ray absorption fine
structure.  The effect of Ru substitution remains confined to the
iron-arsenide layer but neither the static disorder nor the Fe-As bond
strength suffers any change for $x \le$ 0.25.  With further Ru substitution the
static disorder increases while the Fe-As bond strength remains
unchanged.  Also, the Ru-As distance ($\sim$2.42 \AA), different from
the Fe-As distance ($\sim$2.39 \AA), does not show any change in its
force constant with the Ru substitution.  These observations suggest
that the SmFe$_{1-x}$Ru$_x$AsO$_{0.85}$F$_{0.15}$ system breaks down
to coexisting local electronic phases on isoelectric substitution in
the active FeAs layer. 

\bigskip

\noindent Journal reference : {\it Supercond. Sci. Technol.} \href{http://iopscience.iop.org/0953-2048/26/6/065005/}{ 26 (2013) 065005 }\\
DOI: \href{http://dx.doi.org/10.1088/0953-2048/26/6/065005}{ 10.1088/0953-2048/26/6/065005}

\end{abstract}


\maketitle

\section{Introduction}
  
The newly discovered iron-based high $T_{c}$ superconductors
\cite{rev_gen,rev_johnston,rev_str1} have a particular layered
structure with electronically active FePn/Ch (Pn = pnictogen; Ch =
chalcogen) layers alternated by spacer layers.  The superconductivity
and magnetism in these materials are strongly dependent on the
thickness of the active layers (e.g., the height of Pn/Ch atoms from
the Fe-plane) \cite{rev_str1}.  Among the iron-based superconductors,
the REFeAsO (RE=rare earth), the so-called 1111 system with well
defined iron-arsenide active layers stacked together with the spacer
layers shows the highest $T_{c}$ \cite{rev_johnston,rev_str1}.
Generally, atomic substitution in either of the stacking layers is
used to control and manipulate superconductivity and other transport
properties (e.g., a partial substitution of O by F) \cite{rev_str1}.
In addition to the control over the superconductivity, atomic
substitution in the active layer also permits to understand the
transport phenomena \cite{sub_REO,sub_FeAs} and develop new structures
through a detailed information on the role of different layers.  In
particular, it is important to have knowledge of the local atomic
correlations and modification introduced by the substituted atoms in
the layered structure topology.
 
X-ray absorption spectroscopy is an atomic site-specific experimental
probe \cite{Konings}, that does not require any long range crystal
symmetry, and hence permits to have a direct access to the local
atomic correlations.  Indeed, x-ray absorption fine structure (EXAFS)
and x-ray absorption near edge structure (XANES) measurements have
been widely exploited to study the layered high T$_{c}$
superconductots \cite{Saini97,Saini2001,Saini2003,SaiBia2005},
including the iron based superconductors
\cite{OyanagiLa,OyanagiSm,Iad09,Joseph09,Joseph10,Joseph11,Iad10Iad11,Iad12a,Iad12b}.
Earlier, we have explored the effect of different spacer layers (RE of
different size) in the 1111-system \cite{Iad09,Joseph09}, and found
that the interlayer atomic order/disorder should be important in these
materials.  We have also studied the effect of charge density varied
by a partial substitution in the REO spacer layers (O by F), revealing
key information on the interlayer atomic correlations and dynamics
\cite{Joseph11}.  Recently we have focussed on the effect of isovalent
Ru (atomic radius 1.34 \AA) substitution directly in the active FeAs
layer in place of Fe (atomic radius 1.26 \AA) of an optimally doped
SmFe$_{1-x}$Ru$_x$AsO$_{0.85}$F$_{0.15}$ \cite{Iad12b}.  The results
have revealed that the local disorder induced by the Ru substitution
is mainly confined to the FeAs layers which are getting thinner and
decoupled from the SmO spacer layers.  The present work is dedicated
to distinctly identify the random static disorder and the bondlength
fluctuations induced by an isoelectric substitution in the
iron-arsenide active layer.  For the purpose, we have performed
temperature dependent As $K$-edge EXAFS measurements on a series of
SmFe$_{1-x}$Ru$_x$AsO$_{0.85}$F$_{0.15}$ ($x$ = 0.0, 0.05, 0.25 and
0.5) samples.  Consistent with earlier work \cite{Iad12b}, the Ru
substituted system is found to be characterized by different Fe-As and
Ru-As distances.  Incidentally, the force constants for these distances
remain unchanged with increasing Ru substitution, indicating distinct
local electronic phases coexisting in the system.  It appears that the
system breaks down to coexisting nanoscale electronic phases due to
isoelectric substitution, having a direct influence on the fundamental
electronic properties of these materials.

\begin{figure}
\includegraphics[width=10.0 cm]{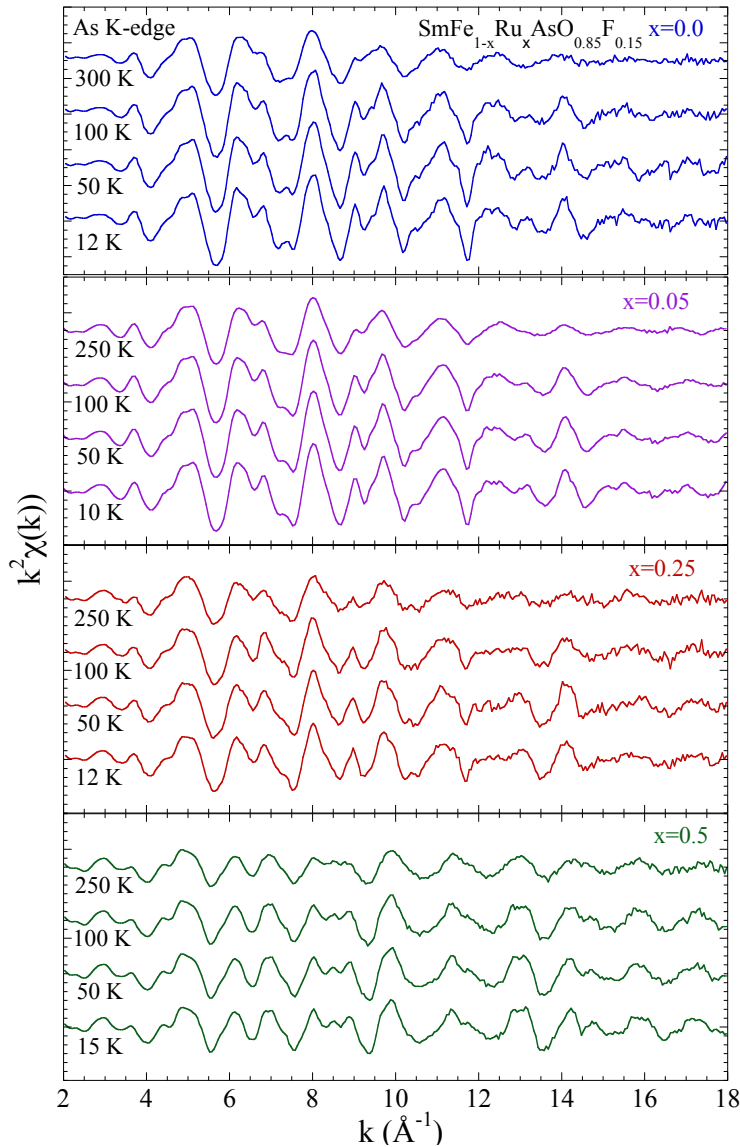}
\caption{\label{fig:1}
Arsenic $K$-edge EXAFS of SmFe$_{1-x}$Ru$_x$AsO$_{0.85}$F$_{0.15}~$
($x$ = 0.0, 0.05, 0.25 and 0.5) at several temperatures (weighted with
k$^2$).  The evolution of the local structure with temperature and Ru
concentration is apparent from the EXAFS oscillations.}
\end{figure}

\section{Experimental details}\label{}

The As $K$-edge (E = 11868 eV) x-ray absorption measurements on powder
samples of SmFe$_{1-x}$Ru$_x$AsO$_{0.85}$F$_{0.15}$ ($x$ = 0.0, 0.25
and 0.5) were performed in transmission mode at the beamline BM26A
\cite{BM26A_ESRF} of the European Synchrotron Radiation Facility
(ESRF), Grenoble.  Measurements on the $x$ = 0.05 sample were carried
out at the XAFS beamline of the ELETTRA, Trieste using similar
experimental approach.  Temperature dependent measurements were
carried out between 10 to 300 K. Several scans were acquired at each
temperature to ensure the spectral reproducibility.  The EXAFS
oscillations were extracted using the standard procedure based on
spline fit to the pre-edge subtracted absorption spectrum
\cite{Konings}.  The superconducting transition temperatures
($T_{c}$), determined by resistivity measurements, are 51 K, 43 K, 14
K and 8 K respectively for the samples with $x$ = 0.0, 0.05, 0.25 and
0.5.  Details on the sample preparation and characterization for
transport and structural properties are described elsewhere
\cite{Tropeano10,SannaPRL11,MartSUST08,RyanPRB10}.

\section{Results and discussions}

\begin{figure}
\includegraphics[width=8.0 cm]{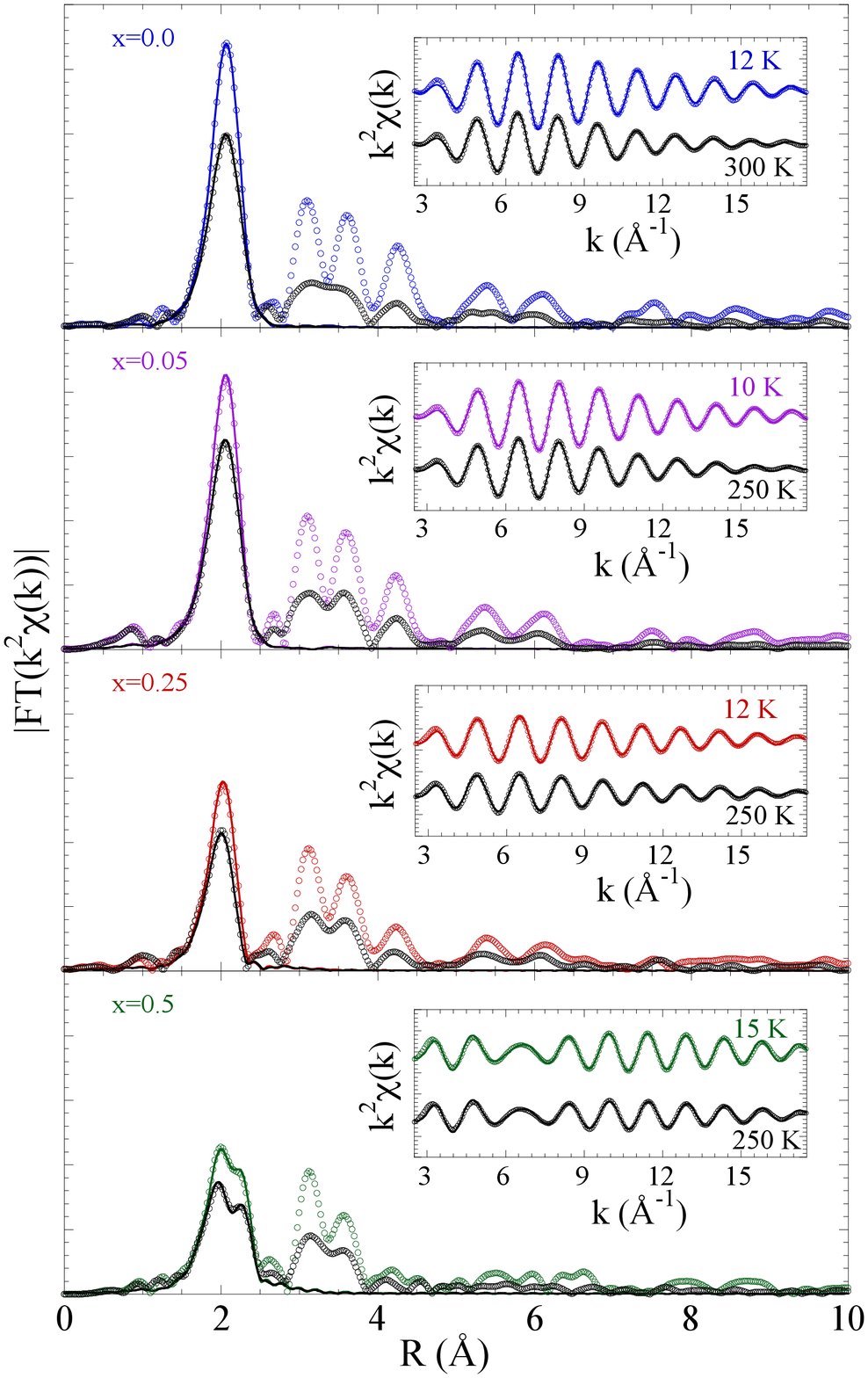}
\caption{\label{fig:2}
Fourier transform magnitudes of the arsenic $K$-edge EXAFS at low and
high temperatures (symbols) together with As-Fe/As-Ru shells model
fits (solid lines) for the SmFe$_{1-x}$Ru$_x$AsO$_{0.85}$F$_{0.15}~$
system.  Insets show the filtered EXAFS oscillations (symbols) and the
corresponding model fits (solid lines).}
\end{figure}

Figure \ref{fig:1} shows k$^2$ weighted arsenic $K$-edge EXAFS of
SmFe$_{1-x}$Ru$_x$AsO$_{0.85}$F$_{0.15}~$ ($x$ = 0.0, 0.05, 0.25 and
0.5) at several temperatures.  The effect of temperature and Ru
substitution is evident from the EXAFS oscillations.  For example,
temperature dependent damping of EXAFS signal can be seen for all the
samples.  Similarly, the effect of Ru substitution can be seen in the
EXAFS oscillations (see, e.g., k $\ge$ 6 \AA$^{-1}$).  These effects can
be better appreciated in the Fourier transforms of the EXAFS,
providing real space information on the partial atomic distribution
around the As atoms.

Figure \ref{fig:2} shows the Fourier transform (FT) magnitudes, obtained using a
Gaussian window (k-range of EXAFS is 3-18 \AA$^{-1}$).  There are four
Fe/Ru near neighbours of arsenic at a distance $\sim$ 2.4 \AA\ (the
main peak at $\sim$ 2 \AA).  The next nearest neighbours of arsenic
are Sm atoms at $\sim$ 3.3 \AA\ and O/F atoms at $\sim$ 3.5 \AA\, follwed by the As atoms at $\sim$ 3.9 \AA\ (the two peaks
between 3-4 \AA), mixed with the multiple scattering contribution due
to Fe/Ru ($\sim$ 4.6 \AA), appearing as FT peak at $\sim$ 4.2 \AA\
(see, e.g. top panel).  It can be seen that the main FT peak is
changing with Ru substitution while other FT peaks due to next near
neighbour atoms suffering much smaller (or negligible) changes.
Indeed, the main peak decreases substantially with Ru and appears as a
clear doublet structure in the $x$ = 0.5 sample.  Similarly, the multiple
scattering Fe/Ru peak at $\sim$ 4.2 \AA\ sustains large change, having
negligible weight in the $x$ = 0.5 sample.  All these data are suggesting
that the atomic disorder due to Ru substitution is mainly confined to
the active layer, consistent with the earlier study \cite{Iad12b}.  
On the other hand, the temperature dependent damping of the FT peaks 
appears almost similar in all the samples.

\begin{figure}
\includegraphics[width=8.5 cm]{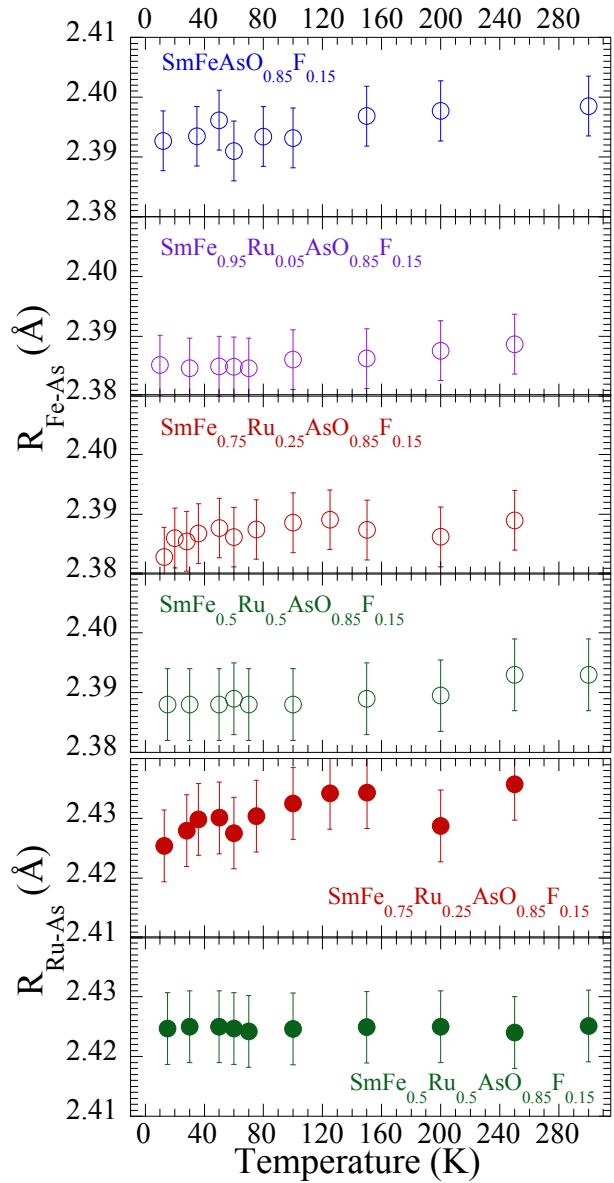}
\caption{\label{fig:3}
Temperature dependence of Fe-As (upper four panels) and Ru-As (lower
two panels) distances for the
SmFe$_{1-x}$Ru$_x$AsO$_{0.85}$F$_{0.15}~$ ($x$ = 0.0, 0.05, 0.25 and
0.5) determined by As K-edge EXAFS analysis.}
\end{figure}

In the single-scattering approximation, the EXAFS is described by the
following general equation\cite{Konings}:
\begin{equation}
\chi(k)= \sum_{i}\frac{N_{i}S_{0}^{2}}{kR_{i}^{2}}f_{i}(k,R_{i})
e^{-\frac{2R_{i}}{\lambda}} e^{-2k^{2}\sigma_{i}^{2}}
sin[2kR_{i}+\delta_{i}(k)]\nonumber
\end{equation}
where N$_{i}$ is the number of neighbouring atoms at a distance
R$_{i}$ from the photoabsorbing atom.  Here, S$_{0}^{2}$ is the
passive electrons amplitude reduction factor, f$_{i}$(k,R$_{i}$) is
the backscattering amplitude, $\lambda$ is the photoelectron mean free
path, $\delta_{i}$ is the phase shift, and $\sigma_{i}^{2}$ is the
correlated Debye-Waller factor measuring the mean square relative
displacement (MSRD) of the photoabsorber-backscatter pairs.

For the As $K$-edge EXAFS in the
SmFe$_{1-x}$Ru$_x$AsO$_{0.85}$F$_{0.15}$, the first shell contribution
involves only the Fe-As/Ru-As bonds, well separated from all other
distant atom contributions \cite{OyanagiLa,OyanagiSm,Joseph11,Iad12b}.
To quantify the temperature dependent atomic displacements we have
analyzed the EXAFS only due to the nearest neighbours.  The filtered
first shell EXAFS are displayed as insets of the Fig.  2.  In the
model fits we have varied the Fe-As/Ru-As distances and the
corresponding $\sigma_{i}^{2}$, while other parameters including the
photo-electron energy origin E$_{0}$ (a value obtained by modeling
five different scans at low temperature), the number of near neighbors
N$_{i}$ (an input from diffraction studies
\cite{Tropeano10,MartSUST08}) and S$_{0}^{2}$ (=1.0) were all kept
fixed for the final iteration.  Phase shifts and amplitude factors
were calculated using the FEFF \cite{Feff}.  The number of independent
data points for this analysis were about 11
(N$_{ind}\sim$(2$\Delta$k$\Delta$R)/$\pi$, where $\Delta$k = 15
\AA$^{-1}$ and $\Delta$R = 1.2 \AA$ $ are the k and R space over which
the data have been analyzed) for a maximum of four parameters fits to
the filtered EXAFS. The k-space (insets) and R-space model fits are
also included in Fig.  \ref{fig:2}.  The errors in the local structural
parameters, determined by the EXAFS analysis, are estimated by
creating correlation maps between different parameters and by analysis
of different scans that are measured in the same conditions, following the
known standards \cite{IXS}.

\begin{figure}
\includegraphics[width=8.0 cm]{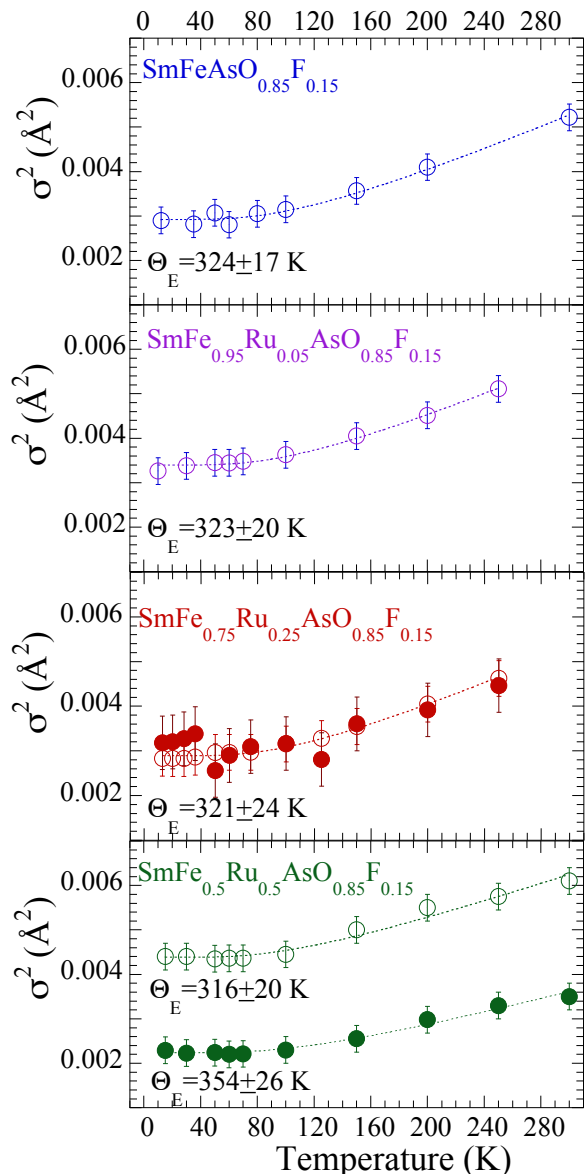}
\caption{\label{fig:4}
Temperature dependence of mean square relative displacements in
SmFe$_{1-x}$Ru$_x$AsO$_{0.85}$F$_{0.15}~$ ($x$ = 0.0, 0.05, 0.25, and
0.5) for the Fe-As (empty symbols) and Ru-As (filled symbols)
bondlengths.  The dotted lines are the correlated Einstein model fits.}
\end{figure}

Figure \ref{fig:3} shows near neighbour (Fe-As and Ru-As) distances as a
function of temperature, obtained from the EXAFS analysis.  The
related $\sigma_{i}^{2}$ are shown in Fig.  \ref{fig:4}.  The Fe-As distance
($\sim$2.39 \AA) is different from the Ru-As distance ($\sim$2.42 \AA)
with the difference ($\sim$0.03 \AA) being smaller than that measured
in isostructural compounds like RuAs-FeAs and RuAs$_{2}$-FeAs$_{2}$
($\sim$0.06 \AA) \cite{Heyding_CJC}.  This suggests the active layer
in SmFe$_{1-x}$Ru$_x$AsO$_{0.85}$F$_{0.15}$ is under chemical pressure
from the REO spacter layer.  Within the experimental uncertainties the
local Fe-As (and Ru-As) distance remains constant with temperature and
Ru concentration, consistent with strongly covalent nature of the
Fe-As (and Ru-As) bonds that is known from earlier studies
\cite{Iad09}.  It is worth recalling that the reported error bars
represent the maximum uncertainty (that is much higher than actual
errors in most of the data points), that are determined by 
considering correlation between the fit parameters, i.e. the 
distances R$_{i}$ and the $\sigma_{i}^{2}$.

The temperature dependence of the $\sigma_{i}^{2}$ permits to
distinctly identify the random static disorder and the dynamic atomic
displacements.  The EXAFS $\sigma_{i}^{2}$ is a sum of temperature
independent ($\sigma_{0}^{2}$) and temperature dependent
$\sigma_{i}^{2}$ (T) terms \cite{Konings}, i.e.,
\begin{equation}
\sigma_{i}^{2} = \sigma_{0}^{2}+\sigma_{i}^{2}(T) \nonumber
\end{equation}
The temperature dependent term can be described by the correlated
Einstein-model \cite{Sevillano,RehrRMP},
\begin{equation}   
\sigma_{i}^{2}(T) = \frac{\hbar}{2 \mu\omega_E} coth(\frac{\hbar\omega_E}{2k_BT}),\nonumber
\end{equation}
where $\mu$ is the reduced mass and $\omega_E$ is the
Einstein-frequency of the pair of atoms involved (i.e.,Fe-As and Ru-As
bonds).  The related Einstein-temperature is
$\Theta_E=\hbar\omega_E/k_B$.  The fits to the correlated Einstein
model are shown as dotted lines in Fig. \ref{fig:4}.  The $\Theta_{E}$ for the
Fe-As bonds are found to be 324$\pm$17 K, 323$\pm$20 K, 321$\pm$24 K
and 316$\pm$20 K for the $x$ = 0.0, 0.05, 0.25 and 0.5 samples
respectively.  The optical phonon modes in SmFeAsO 
are 201 cm$^{-1}$ (A$_{1g}$ involving As atom displacements) and 208
cm$^{-1}$ (B$_{1g}$ involving Fe atom displacements) \cite{Dore08,Chu08}.  
These
frequencies are quite similar to the EXAFS findings, the Einstein
frequency of the Fe-As to be about 225 cm$^{-1}$ ($\Theta_{E}\sim$324
K).  The $\Theta_{E}$ for the Fe-As bonds are lower than the
$\Theta_{E}$ for the Ru-As bonds.  The $\Theta_{E}$ for the Ru-As
bonds are found to be 353$\pm$42 K and 354$\pm$26 K respectively for
the $x$=0.25 and $x$=0.5 samples.  The $\sigma_{0}^{2}$ representing
the random static disorder, is significantly higher for the $x$=0.5
sample, $\sim$ 0.0026 \AA$^2$, compared to $\sim$0.0006 \AA$^2$ for
$x$=0.0, 0.05 and 0.25 samples.  Similar static disorder and force
constants (5.09, 5.06 and 4.99 eV/\AA$^{2}$ respectively) for the Fe-As
bonds in $x$ = 0.0, 0.05 and 0.25 indicate that the atomic disorder in
the FeAs layer may not have a direct effect the electronic
transport.  Indeed, the residual resistivity $\rho_{0}$ changes
anomalously as a function of Ru concentration.  The $\rho_{0}$ is
$\sim$0.3 m$\Omega$ cm for the pure SmFeAsO$_{0.85}$F$_{0.15}$ system
and increases sharply with the Ru substitution reaching a maximum
value of about 2 m$\Omega$ cm for $x$ = 0.25 while the T$_{c}$ decreases
\cite{Tropeano10}.  With further Ru substitution the $\rho_{0}$
decreases by half at $x$ = 0.5 ($\rho_{0}\sim$2 m$\Omega$ cm) and has a
similar value of $\rho_{0}\sim$0.3 m$\Omega$ cm for the $x$ = 1.
Therefore, it appears that the impurity scattering from the
substituted ruthenium being dominant to describe the transport
phenomena for $x\le$0.25.

On the other hand, the larger static disorder of the Fe-As bonds in
$x$ = 0.5 sample ($\sigma_{0}^{2}\sim$0.0026 \AA$^2$), albeit with the
force constant being similar to the other samples (4.83 eV/\AA$^{2}$),
suggests that some different mechanism should be active to describe
the electronic transport, i.e. decreased $\rho_{0}$ and the $T_{c}$
for $x\ge$ 0.25.  In addition, the force constant for the Ru-As
distance remains the same for $x$ = 0.25 and 0.5 samples, (6.04 and 6.07 eV/\AA$^{2}$, respectively).  Therefore, it is likely that the title 
system is phase separated even at $x$ = 0.25, and the reduced $\rho_{0}$ 
from $x$ = 0.25 to $x$ = 0.5 is merely due to the increased density of states
with increasing Ru because of more extended Ru 4d states than Fe 3d,
consistent with the density functional theory calculations
\cite{Tropeano10}.  It should be mentioned that the active and spacer
layers are getting decoupled and thinner with the Ru substitution
\cite{Iad12b}, i.e., electronically the system contains active FeAs
layers which have poorer screening from the spacer layers, and hence
can suffer phase separation as the case of ternary
FeSe$_{1-x}$Te$_{x}$ \cite{Joseph10}.  Since the characteristic length
scale of EXAFS is about a nanometer, it can be fairly argued that the
observed phase separation is at a nanometer length scale, that is
consistent with $\mu$SR measurements \cite{SannaPRL11} on the similar
system.  It is also interesting to note that similar nanoscale
textures have been observed in the Ru-substituted BaFe$_{2}$As$_{2}$
(122) system by NMR experiments due to inhomogeneous destruction of
antiferromagnetic order by Ru substitution \cite{NMR122}.

We can also notice that the $\sigma_{i}^{2}$ for the Ru-As bondlength
manifests an upturn at a temperature $\sim$ 40 K (Fig.  4).  This
appears to be consistent with the anomaly observed in zero field
$\mu$SR measurements, sensitive to short range magnetic order,
revealing similar change in the Ru substituted 1111-system
\cite{SannaPRL11}.  However, the observed change is very small to be
stressed further and more experimental work is needed before it can be
argued if the short range magnetic order is coupled to the charge and
atomic displacements for particular Ru concentrations.

In summary, we have studied temperature dependent local structure of
SmFe$_{1-x}$Ru$_x$AsO$_{0.85}$F$_{0.15}$ system for varying Ru
concentration by As $K$-edge EXAFS. We find that Ru substitution
effect is mainly confined to the electronically active FeAs layer, and
the Ru substituted system has different Fe-As and Ru-As distances.
The force constants of the Fe-As and Ru-As bonds do not show any
change with the Ru concentration, indicating coexisting electronic
phases in the isoelectronic substitution.  The static disorder in the
Fe-As bonds remains unchanged as the bond strength for $x \le$ 0.25,
suggesting that the transport properties of the system should be
described mainly by the impurity scattering in FeAs layers.  With
further Ru substitution, the extended Ru 4d states affects
substantially the electronic density of states at the Fermi level, and
hence the transport phenomena. On the basis of present results we can
conclude that upon isoelectric substitution the title system breaks
down into coexisting nanoscale electronic phases due to the
frustration of interlayer atomic correlations.

\section*{Acknowledgments}
The authors thank Sergey Nikitenko and Miguel Silveria of BM26A, ESRF,
Grenoble and Luca Olivi, Giuliana Aquilanti and Nicola Novello, of XAFS,
ELETTRA, Trieste, for their active cooperation in the EXAFS
measurements.


\end{document}